# Realization of a Contactless Acoustic Levitation Motor via Doublet Mode Control and Autoresonance

Solomon Davis, Izhak Bucher

**Abstract**

This paper demonstrates analytically and experimentally an acoustic levitation motor which has the ability to levitate and rotate an object in the air without mechanical contact. To realize such a device two core methods are applied simultaneously; (i) resonance tracking with an Autoresonance feedback loop, (ii) generation of controlled structural traveling waves. The purpose of the first method is to achieve near-field acoustic levitation, which can levitate an object of a few kilograms. In this research, this is accomplished through high amplitude vibration of an aluminum annulus at ultrasonic frequencies (~30kHz). For high efficiency, the annulus is designed to have a very high Q value, and operating even slightly off resonance ceases levitation. Compounding this is the fact that the natural frequency constantly drifts as ambient conditions and loading change. To accommodate such a drift, and produce stable levitation automatically, a resonance tracking feedback loop is employed here. Simultaneously, the purpose of the second method is to achieve propulsion forces on the levitated object by propagating and controlling traveling waves in the aluminum annulus to create a thin layer of rotating air beneath the levitated body. Even though a single vibration mode can produce only standing waves, an axisymmetric structure possesses two modes per natural frequency, and excitation of a doublet mode pair can generate effective travelling waves. The present paper develops the theory behind the use of the Autoresonance feedback method for achieving constant levitation and propagating travelling waves in co- and counter rotating directions. It will be shown that all this can be accomplished with only single sensor. The result is a stable, repeatable and a highly controllable contactless acoustic levitation motor.

1. Introduction

An acoustic levitation motor [1] is a device that uses near-field acoustic levitation [2, 3] and structural traveling waves to simultaneously (i) levitate and object and (ii) apply a torque on the object to rotate it. Like conventional motors, this device consists of two main components, the rotor and the stator. However, in the case of the acoustic motor, the rotor is the acoustically levitated object, and does not experience any mechanical contact. The advantage of such a device is the ability to position objects in space which are sensitive to mechanical contact, such as a silicon wafer in a metrology lab. In previous works such as in [4], a functioning acoustic motor was realized, and the angular position of the levitated rotor could be controlled in closed loop without contact. Though not reported, a pressing problem with this device was the probable lack of stable acoustic levitation, due to frequency drift of the stator. Because of this drift, which is apparent in all near-field acoustic levitation systems [5, 6], it's likely that levitation was frequently lost, and that performance of the motor was neither stable nor repeatable, requiring constant manual tuning. Therefore, it is the main goal of this research to improve upon [4], and stabilize the levitation of the rotor by adding resonance frequency tracking feedback loop, while retaining the ability to generate high-quality traveling waves. The specific frequency tracking algorithm chosen was Autoresonance [7, 8]. Autoresonance is well known in ultrasonic applications, but using it for control of traveling waves is novel and requires a special formulation and method.

When two planar surfaces in close proximity and surrounded by a compressible fluid experience a relative oscillatory motion, a layer of compressible fluid becomes entrapped in the gap between the surfaces. Because this layer is compressed, its average pressure becomes higher than the ambient pressure, and the layer is a "squeeze film". Thus, a net repulsing force is applied on the surfaces [9, 10, 6]. This net force can oppose gravity and levitation can occur. The phenomenon is referred to as near-field acoustic levitation [2, 3] and has the ability to levitate a planar object of a few kilograms 5 to 200 microns above the vibrating surface.

In addition to near field acoustic levitation, which only has the ability to apply a normal force, it was shown in earlier works [2, 3, 11], that travelling flexural waves could induce a pressure gradient in the wave's direction. This gradient will cause the squeeze film to flow, and a shear force acts on the levitated object. The method was demonstrated in [12, 13] where the position of a levitated object was controlled along guide rails. Unlike a simple beam which cannot experience high quality traveling waves throughout its entire length [14], it was shown in [15] that an annulus can experience consistent and high quality traveling waves throughout the structure. In [16] it was shown that to achieve a high amplitude travelling wave, it's necessary to maintain the cyclic symmetry of the annulus. Therefore, if the actuators are placed in an equally spaced distribution, modal contamination can be reduced.

In [4] Gabai et al demonstrated a functioning acoustic motor by simultaneously levitating and rotating a flat disk above an annulus. The control methodology they devised first plotted the traveling wave response of the annulus over a wide range of actuator inputs. This plot was named the "Travelling Wave Ratio Map". Next, this map was used as a guide for manipulating the torque on the levitated rotor, as one could clearly see which inputs produced the desired torque. It can be seen from their results that they were indeed successful in controlling the angular position of the rotor. However, their results and methodology are not viewed as a complete realization of a functioning and stable acoustic motor that operates without frequent manual tuning.

The conclusion that the previous work was incomplete is based on the following observations. (i) Unstable Levitation: The natural frequency of high-Q mechanical oscillators tend to drift as ambient conditions change [17], as was certainly the case with the acoustic levitation device [5]. Typically a Phase-locked Loop is used to lock onto resonance with such a system [18, 19]. However in [4], the annulus was excited at a constant frequency. This makes it unlikely that they were able to achieve stable levitation [5]. (ii) Lack of Validation: They did not sufficiently demonstrate that an experimental Travelling Wave Ratio Map could be generated and used for control. Only theoretical maps were provided.

To improve upon the work of [4], various steps will be taken such as the simplification and validation of the annulus's model. For example, it will be assumed that the annulus is well-machined and nearly axisymmetric. If true, then the control of travelling waves in the annulus is greatly simplified. The most essential improvement however, will be the addition of a resonance frequency tracking feedback loop. This is to ensure that the motor will operate in a stable and efficient manner at all times, regardless of any frequency drift.

A number for frequency tracking algorithms exist such as a Phase-locked loop [19] or Optimum Seeking (Hill Climbing) [5], and any of these methods could have been used. The algorithm chosen in this work was Autoresonance (AR), also known as Self Excitation [7, 8]. This method has found widespread use in Auto-tuning of PID controllers [20] and automatic excitation of quartz tuning forks [21]. In this research, the reason it was selected over the other algorithms was its simplicity, it requires no PID or controller tuning, and the speed with which it locks onto resonance. It should be noted however, that applying of such a feedback loop, and also decoupling it from traveling wave control can be difficult. Therefore, it is the goal of this paper to develop the mathematics and methodology to achieve constant and stable levitation of the rotor, even as the torque on the rotor is manipulated. This paper consists of two main parts, the theoretical and the experimental sections.

The theoretical section begins by describing the physical system's dimension and design considerations. Next, a mathematical model of the annulus will be presented which will be used throughout. After this, it will be shown how to decompose the steady state response of the annulus into its traveling and standing wave components. Following this, a relationship between the spatial forces and the modal forces is presented, and with this, a method for manipulating the traveling wave component of the annulus' vibration response is proposed. Next, the AR feedback loop is described and a method for decoupling the loop from travelling wave control is developed.

The experimental section begins by manipulating the modal response of the annulus in open loop to validate the model and proposed traveling wave control method. Next, the annulus is excited by AR to determine if the selected modes can still be controlled in closed loop. Finally, a functioning acoustic levitation motor is demonstrated by controlling its steady state angular speed and direction. These data are also compared with the vibration response of the annulus to determine if the motor behaves as the theory predicts.

2. **Motor Design**

The device used in this paper consists of four main components. The actuators, which provide the source of vibration, the annulus which produces a rotating or stationary squeeze film, the squeeze film itself, and the rotor, typically a flat disk which sits atop the film. Three piezoelectric actuators (model FBL28452HS with natural frequency 28 kHz) spaced 120 degrees apart are connected to the bottom of the annulus. The annulus is an aluminum tapered ring with an inner diameter of 100 mm, outer diameter of 150 mm, inner thickness of 2.5 mm and outer thickness of 5 mm. Vibrating the annulus with high frequency ultrasonic vibration near 28 kHz will generate a squeeze film and the rotor will levitate. A diagram of the annulus with the piezoelectric actuators can been seen in Fig.1.

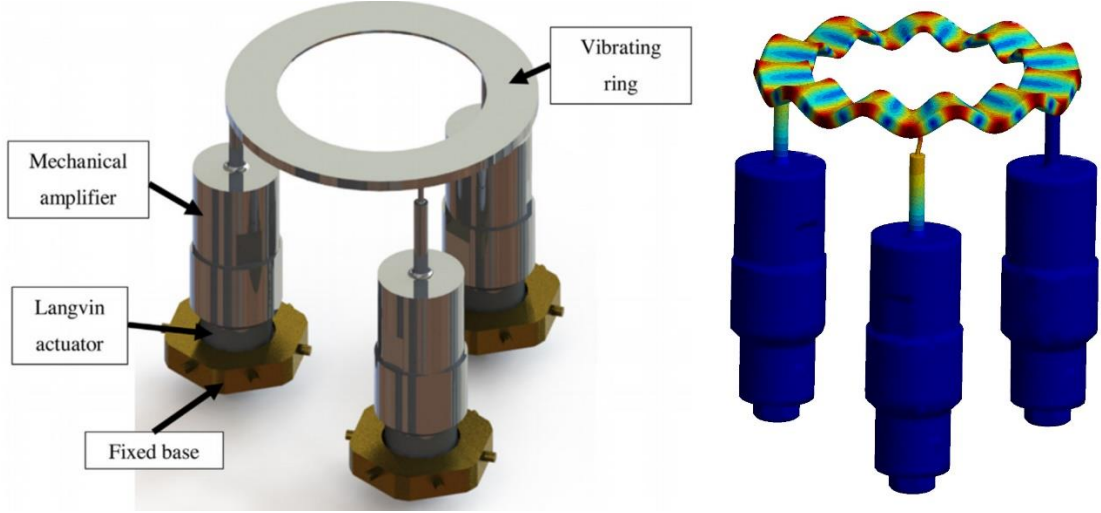

Figure 1. (Left) The annulus (stator) with three piezoelectric actuators spaced 120 degrees apart [4]. (Right) Finite element model showing the deformed shape at the designated mode of vibration.

### 3. Annulus Model

Due to the near axial symmetry, the relevant eigenfunctions of the annulus occur in pairs for each natural frequency. These doublet modes will have similar spatial response, but rotated in space with respect to each other. In general, the $\kappa th$ doublet mode pair corresponding to the same natural frequency $\omega_\kappa$ can be written as [4]

$$\begin{aligned}\phi_{c,\kappa}(r,\theta) &= A_\kappa \cos(\kappa\theta)R(r) \\ \phi_{s,\kappa}(r,\theta) &= A_\kappa \sin(\kappa\theta)R(r)\end{aligned} \quad (1)$$

for amplitude $A_\kappa$, wave number $\kappa$, angular location $\theta$ and radial location $r$. The term $R(r)$ describes the radial dependency of the deflection, but for simplicity it will be assumed that that $R(r)=1$. $\phi_{c,\kappa}$ and $\phi_{s,\kappa}$ are referred to as the cosine and sine modes. By manipulating the response of both these modes simultaneously, a structural travelling wave can be propagated through the annulus.

To achieve efficient vibration amplitude, the annulus was designed so that one of the doublet mode pairs has a natural frequency at 28 kHz, the same natural frequency as the actuators. These modes are named the "levitation modes" and for this system have a wave number of $\kappa=11$, and natural frequency $\omega_{11}$.

The complete steady-state response of the axisymmetric annulus, which is uniform in the radial direction by design (see [3]), in the frequency domain is

$$U(\theta,s) = \phi_c(\theta)\eta_c(s) + \phi_s(\theta)\eta_s(s) + \sum_{\kappa=1\notin 11}^{\infty}\left[\phi_{c,\kappa}(\theta)\eta_{c,\kappa}(s) + \phi_{s,\kappa}(\theta)\eta_{s,\kappa}(s)\right], \quad (2)$$

where $\phi_c \triangleq \phi_{c,11}$, $\phi_s \triangleq \phi_{s,11}$ are the levitation modes, $\eta_c \triangleq \eta_{c,11}$, $\eta_s \triangleq \eta_{s,11}$ are the Laplace transformed modal coordinates corresponding to the levitation modes, $\phi_{c,\kappa\notin 11}$, $\phi_{s,\kappa\notin 11}$ are the other modes, and $\eta_{c,\kappa\notin 11}$, $\eta_{s,\kappa\notin 11}$ are the other modal coordinates. Also, $s$ is the Laplace variable. A modal coordinate can be expanded as [22]

$$\eta_\kappa(s) = Q_\kappa G_\kappa(s) = \frac{Q_\kappa}{s^2 + 2\zeta_\kappa \omega_\kappa + \omega_\kappa^2}, \tag{3}$$

for modal force $Q_\kappa$, modal transfer function $G_\kappa$, modal damping $\zeta_\kappa$ and natural frequency $\omega_\kappa$. Inserting (3) into (2) one obtains

$$U(\theta,s) = G_{11}(s)\left[\phi_c(\theta)Q_c + \phi_s(\theta)Q_s\right] + \sum_{\kappa=1\notin 11}^{\infty} G_\kappa(s)\left[\phi_{c,\kappa}(\theta)Q_{c,\kappa} + \phi_{s,\kappa}(\theta)Q_{s,\kappa}\right], \tag{4}$$

for the most influential levitation modal forces $Q_c \triangleq Q_{c,11}$ and $Q_s \triangleq Q_{s,11}$. When the excitation frequency is near the natural frequency of the levitation modes $\omega_{11}$, the annulus response is practically a truncated response

$$\begin{aligned}U(\theta,i\omega)\big|_{\omega\approx\omega_{11}} &\approx U_{tr}(\theta,i\omega) = \phi_c(\theta)\eta_c(i\omega) + \phi_s(\theta)\eta_s(i\omega)\\ &= G_{11}(i\omega)\left[\phi_c(\theta)Q_c + \phi_s(\theta)Q_s\right]\end{aligned}. \tag{5}$$

4. Traveling and Standing Wave Components and Torque

In reality the response of the annulus will contain the superposition of traveling and standing waves. This section demonstrates how to decompose the annulus response into its traveling and standing components so that the torque on the levitated rotor can be estimated from the travelling wave component.

If the annulus is excited near the natural frequency of the levitation modes by $\omega \approx \omega_{11}$, then the truncated response is a good approximation of the total response. Writing the modal amplitudes in their complex form

$$U_{tr}(\theta,i\omega) = (a_c + ib_c)\cos(\kappa\theta) + (a_s + ib_s)\sin(\kappa\theta), \tag{6}$$

for a set of parameters $a_c$, $b_c$, $a_s$ and $b_s$. This response can be decomposed such that [23]

$$U = U_T + U_S, \tag{7}$$

where $U_T$ and $U_S$ are the traveling and standing wave components of the response. An alternative representation of (7) is [24]

$$U_{tr}(\theta, i\omega) = c_+ e^{i\kappa\theta} + c_- e^{-i\kappa\theta}$$
$$= U_{tr} = |c_+ + c_-| e^{i\varphi_a} \cos\kappa\theta + |c_- - c_+| e^{i\varphi_b} \sin\kappa\theta, \quad . \tag{8}$$
$$\varphi_b(t) = \arg[c_- - c_+], \quad \varphi_a(t) = \arg[c_+ + c_-]$$

Then the amplitude and direction of the travelling wave component is represented by the amplitude and sign of the parameter $A_T$, where

$$A_T = |c_+| - |c_-| . \tag{9}$$

Now that the traveling wave component has been extracted from the response, we'd like to know how it influences the torque on the levitated object. The relation between the applied torque, air pressure and air gap is [25]

$$T \approx -\int_0^{2\pi}\int_{r_{in}}^{r_{out}} \frac{hr}{2} \frac{\partial p}{\partial \theta} dr d\theta , \tag{10}$$

where $h$ is the air gap, $r$ is the radial coordinate, $p$ is the pressure and $r_{in}$ and $r_{out}$ are the inner and outer radii of the annulus. This relation implies only a travelling wave can induce a torque on the levitated rotor [25], and that the strength and direction of the torque is directly related to the amplitude and direction of the travelling wave.

By (10) it's hypothesized that the amplitude and direction of the traveling wave is proportional to the strength and direction the torque acting on the rotor.

$$A_T \propto T \tag{11}$$

In the experimental section, $A_T$ will be extracted from vibration of the annulus and compared with $T$ so that (11) may be validated.

5. **Derivation of the Modal Forces**

In the previous section it was hypothesized that the traveling wave response of the annulus plays a direct role in the applied toque on the levitated rotor. In this section the transformation relating the spatial forces to modal forces for an axisymmetric annulus is presented. Following this, a method for manipulating the traveling wave component of the response $A_T$ is developed.

Through the Principle of Virtual Work [26], Gabai et al [4] developed a transformation $\mathbf{T}$ to convert the spatial forces of the actuators $f_i(t)$ to generalized modal forces $q_i(t)$. In this derivation the rigid body response $\eta_0$ was considered, as well as the response from levitation modes $\eta_c$ and $\eta_s$. However, in the present paper it's assumed that the excitation frequency is near the levitation mode's natural frequency, $\omega \approx \omega_{11}$, and the other modal responses, including

$\eta_0$ are negligible. Therefore, the rigid body modal force $q_0$ is omitted from the transformation. The spatial distribution of the forces on the annulus spaced 120 degrees apart is

$$f(\theta) = f_1 \delta(\theta) + f_2 \delta\left(\theta - \frac{2\pi}{3}\right) + f_3 \delta\left(\theta - \frac{4\pi}{3}\right) \tag{12}$$

Then through virtual work, the transformation $\mathbf{T}$ is defined as [4]

$$\begin{bmatrix} q_c \\ q_s \end{bmatrix} = \underbrace{\begin{bmatrix} A_\kappa & A_\kappa \cos\left(\frac{2\pi\kappa}{3}\right) & A_\kappa \cos\left(\frac{4\pi\kappa}{3}\right) \\ 0 & A_\kappa \sin\left(\frac{2\pi\kappa}{3}\right) & A_\kappa \sin\left(\frac{4\pi\kappa}{3}\right) \end{bmatrix}}_{\mathbf{T}} \begin{bmatrix} f_1 \\ f_2 \\ f_3 \end{bmatrix}. \tag{13}$$

Assuming a single excitation frequency, (13) can be expressed in terms of complex amplitudes, where $F_i = \mathcal{F}(f_i)$ and $Q_i = \mathcal{F}(q_i)$.

$$\begin{bmatrix} Q_c \\ Q_s \end{bmatrix} = \underbrace{\begin{bmatrix} A_\kappa & A_\kappa \cos\left(\frac{2\pi\kappa}{3}\right) & A_\kappa \cos\left(\frac{4\pi\kappa}{3}\right) \\ 0 & A_\kappa \sin\left(\frac{2\pi\kappa}{3}\right) & A_\kappa \sin\left(\frac{4\pi\kappa}{3}\right) \end{bmatrix}}_{\mathbf{T}} \begin{bmatrix} F_1 \\ F_2 \\ F_3 \end{bmatrix}. \tag{14}$$

And the spatial force and modal force vectors

$$F_{act} = \begin{bmatrix} F_1 \\ F_2 \\ F_3 \end{bmatrix}, \quad Q = \begin{bmatrix} Q_c \\ Q_s \end{bmatrix}. \tag{15}$$

Now that a relationship between the spatial forces $F_{act}$ and the modal forces $Q$ has been developed, a strategy for manipulating the modal forces and hence the modal responses is proposed.

The strategy will be to generate three forces by the three actuators at the same amplitude and frequency but with varying relative phases. The first actuator will be the reference actuator and the phases of the other two will be equal but opposite. This scheme was first suggested in [4].

$$F_{act}(\phi_{act}) = \begin{bmatrix} 1 \\ e^{i\phi_{act}} \\ e^{-i\phi_{act}} \end{bmatrix} F = b(\phi_{act}) F \tag{16}$$

To see its effect on $Q$ for $\kappa = 11$, transform $F_{act}$ with $\mathbf{T}$, apply trigonometric identities and Euler's Formula,

$$\mathbf{T}F_{act} = \mathbf{T}b(\phi_{act})F = \begin{bmatrix} Q_c \\ Q_s \end{bmatrix} = \begin{bmatrix} 1-\cos(\phi_{act}) \\ i\sqrt{3}\sin(\phi_{act}) \end{bmatrix} FA, \qquad (17)$$

where $A \triangleq A_{11}$. Since the three forces are a function of the reference force $F$ and $\phi_{act}$, two transfer functions relating the reference force to the two modal forces are developed. Taking (17) and dividing by $F$, one can obtains these transfer functions

$$\Gamma_c(\phi_{act}) = \frac{Q_c}{F} = \mathbf{T}_1 b(\phi_{act}) = A(1-\cos(\phi_{act}))$$
$$\Gamma_s(\phi_{act}) = \frac{Q_s}{F} = \mathbf{T}_2 b(\phi_{act}) = i A\sqrt{3}\sin(\phi_{act}) \qquad (18)$$

By varying $\phi_{act}$ by 360 degrees, one can visualize the amplitude and phase of $\Gamma_c$ and $\Gamma_s$, the modal forces per reference force $F$, as seen in Fig.2.

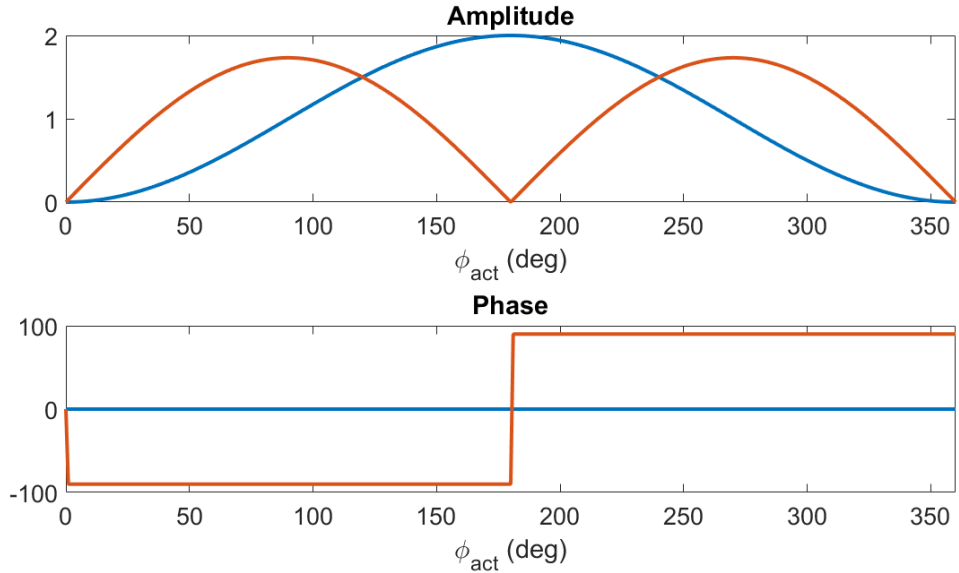

Figure 2. Modal force response plots $\Gamma_c$ and $\Gamma_s$, amplitude and phase when varying $\phi_{act}$, $A=1$.

To visualize how (18) influences the traveling wave response, consider the truncated response (5) and insert (18)

$$U_{tr}(\theta, i\omega) = AFG_{11}(i\omega)\left[\Gamma_c(\phi_{act})\cos(\kappa\theta) + \Gamma_s(\phi_{act})\sin(\kappa\theta)\right]. \qquad (19)$$

For a constant frequency $\omega$, $AFG_{11}(i\omega)$ is some complex constant. By setting $AFG_{11}(i\omega) = 1$, and decomposing (19) as in (9), we are able to show an example plot for $A_T(\phi_{act})$ vs $\phi_{act}$, where $A_T$ was calculated by (9). This can be seen in Fig.3.

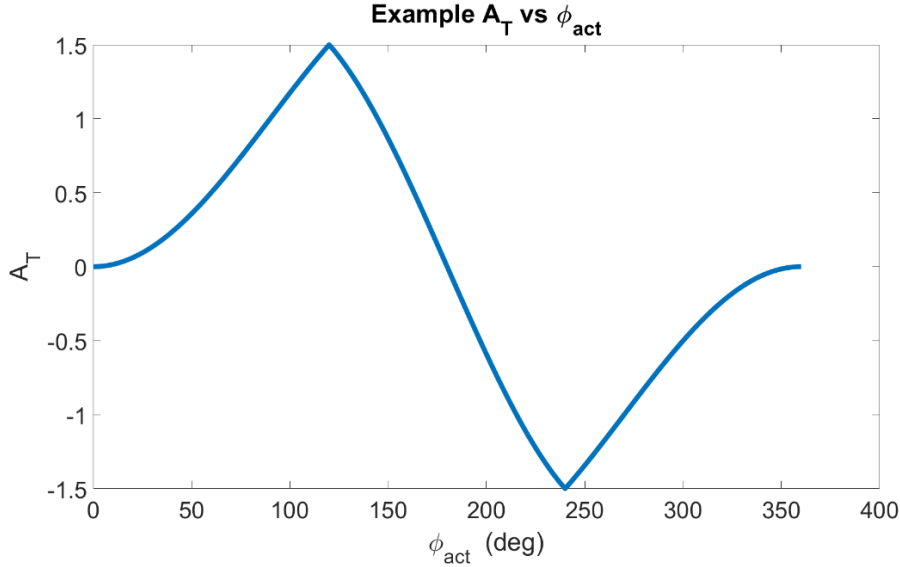

Figure 3. Example of the traveling wave amplitude and direction $A_T$ vs $\phi_{act}$, for complex constant $AFG_{11}(i\omega)=1$, where $A_T$ was calculated by (9).

From Fig.3 it's observed that a monotonic region of $A_T$ exists, from about 120 to 240. This monotonic region will be useful for control of the torque applied on the rotor and will later be observed in the experimental data.

### 5.1 The Nearly Axisymmetric Model and the Angular Orientation of Actuator Forces to Vibration Modes

In the previous section, the assumed relative angle between the reference actuator and the modes was zero, as manifested in (12). But in reality, this is not a valid assumption for a physical system. For example, consider a perfect annulus, where the modes have no set location. Then adding even an infinitesimal imperfection to the model will have a great influence on the modes' location, even though it's influence on parameters like mode shape, modal amplitude and natural frequency will be negligible. Therefore, the effect of any imperfections on the location of the modes cannot be ignored and must be considered in the model. This is why the relative angle between the reference actuator and the modes cannot be assumed to be zero. This leads to the declared definition of a nearly axisymmetric annulus; an annulus whose doublet modes have identical amplitude and natural frequency (1), (5), but with a fixed location in space.

For most systems it's not possible to excite vibration modes in a specific manner if the orientation of the actuators to the modes is unknown. However, for the case of the nearly axisymmetric annulus, it will be shown that the vibration response of the annulus does not depend on the modes' actual angular orientation. This makes control of the annulus' response surprisingly simple, as the physical orientation of the modes need not be considered.

In general, the applied forces (12) should be written as

$$f(\theta) = f_1\delta(\theta-\theta_0) + f_2\delta\left(\theta - \frac{2\pi}{3} - \theta_0\right) + f_3\delta\left(\theta - \frac{4\pi}{3} - \theta_0\right), \tag{20}$$

where $\theta_0$ is the absolute location of the actuators in the doublet modes' spatial coordinate system. In this case the transformation $\mathbf{T}$ becomes

$$\mathbf{T}_{\theta_0} = \begin{bmatrix} A\cos(\theta_0) & A\cos\left(\frac{2\pi\kappa}{3} + \theta_0\right) & A\cos\left(\frac{4\pi\kappa}{3} + \theta_0\right) \\ A\sin(\theta_0) & A\sin\left(\frac{2\pi\kappa}{3} + \theta_0\right) & A\sin\left(\frac{4\pi\kappa}{3} + \theta_0\right) \end{bmatrix}. \tag{21}$$

At this point (17) appears to no longer hold and the modal response plot may be warped. However, it will be shown that response is still manageable.

First, calculate the response when $\theta_0 \neq 0$. The force distribution is (20), and the transformation to modal forces is (21). Recalculating the response as in (17) for input force vector (16), but this time using $\mathbf{T}_{\theta_0}$, one obtains

$$\begin{aligned} U_{tr}(\theta) &= A^2 FG_{11}\cos(\kappa\theta)\left[\cos(\theta_0) + \cos\left(\theta_0 + \frac{2\pi\kappa}{3}\right)e^{i\phi_{act}} + \cos\left(\theta_0 + \frac{4\pi\kappa}{3}\right)e^{-i\phi_{act}}\right] \\ &+ A^2 FG_{11}\sin(\kappa\theta)\left[\sin(\theta_0) + \sin\left(\theta_0 + \frac{2\pi\kappa}{3}\right)e^{i\phi_{act}} + \sin\left(\theta_0 + \frac{4\pi\kappa}{3}\right)e^{-i\phi_{act}}\right] \end{aligned}. \tag{22}$$

Using trigonometric identities and Euler's Formula and relation (18), (22) reduces to

$$U_{tr}(\theta) = AFG_{11}\left[\Gamma_c \cos(\kappa\theta - \theta_0) + \Gamma_s \sin(\kappa\theta - \theta_0)\right]. \tag{23}$$

In comparing (19) with (23) it's observed for $\theta_0 \neq 0$, that the effective modal response follows the location of the reference actuator, it is also rotated by $\theta_0$. Because the effective modes of the annulus (which by (22) are combinations of the sine and cosine modes) are apparently always aligned with the reference actuator, the effective relative angle between the reference actuator and the modes is always zero and therefore (14) is always valid. This is a particularly useful result because controlling the annulus does not require one to find the physical orientation of the modes ahead of time. The response will be the same regardless of their orientation with the actuators, and is solely dependent on the location of the chosen reference actuator.

### 6. Autoresonance, Modal Filtering and Travelling Wave Control

The analysis from the previous sections assumed that sufficient vibration amplitude is present in the annulus to achieve acoustic levitation of the rotor [27]. In practice, this can only be achieved when the annulus is vibrating near the frequency $\omega_{11}$.

Because this frequency drifts as ambient conditions and loading change, an Autoresonance feedback loop is added to stabilize the rotor levitation. Like a phase-locked loop (PLL), AR operates by exciting a frequency that produces a specific input-output phase shift by the plant, and this phase shift is chosen by the operator. For example, to lock onto a natural frequency, the phase shift is chosen to be $-\pi/2$. However, one drawback of the algorithm is apparent when the plant possesses multiple vibration modes. In this case multiple frequencies will produce identical phase shifts. The algorithm will therefore automatically excite one of these frequencies [8], but the specific one cannot be chosen by the operator. Another potential problem is if the natural frequency does not produce a constant phase shift. This can be the case when doublet modes exist. These two obstacles are indeed present with the annulus, and two modifications must be added to the AR algorithm to ensure proper resonance tracking at all times. These modifications are a Modal Filter (MF) [28], to ensure $\omega_{11}$ produces a constant phase shift as $\phi_{act}$ is manipulated, and a bandpass filter, to ensure $\omega_{11}$ is excited, even if other frequencies produce the same phase shift [29].

The first step in analyzing the behavior of the annulus excited by AR is to consider only the truncated response of the annulus (19). From this we can determine if it's even possible to excite the annulus at $\omega_{11}$ with AR. Rewriting (19) with the eigenfunctions expressed symbolically,

$$U_{tr}(\theta,\phi_{act},s) = FG_{11}(s)\left[\phi_c(\theta)\Gamma_c(\phi_{act}) + \phi_s(\theta)\Gamma_s(\phi_{act})\right] . \tag{24}$$

Eq. (24) is the feedback signal of the AR loop at sensor location $\theta$. Dividing (24) by $F$ produces the transfer function of the annulus with respect to the reference force. This is the effective plant excited by AR.

$$G_{tr}(\theta,\phi_{act},s) = \frac{U_{tr}}{F} = G_{11}(s)\left[\phi_c(\theta)\Gamma_c(\phi_{act}) + \phi_s(\theta)\Gamma_s(\phi_{act})\right] \tag{25}$$

To automatically excite the system at its natural frequency $\omega_{11}$, $G_{tr}$ must be enclosed in an AR feedback loop. But to first understand how AR works, consider an AR loop enclosed on the general second order plant $L$. In its simplest form, an AR loop consists of a nonlinear amplitude scaling element $\Psi$ and a phase shifting element $P$. Because the loop contains nonlinear elements, the undamped oscillation excited in the plant can exhibit a limit cycle. A diagram of the AR feedback loop can be seen in Fig.4.

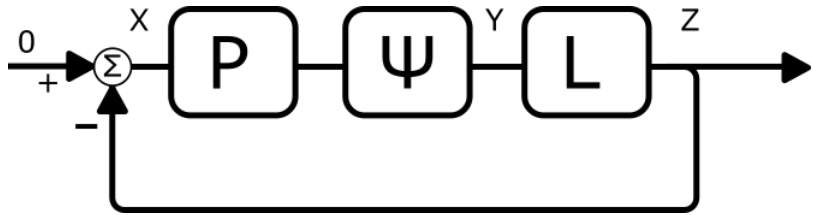

Figure 4. Basic AR loop [8], exciting the plant $L$ with a nonlinear scaling element $\Psi$ and phase shifting digital filter *P.*

For a feedback loop of Fig.4, the frequency of the limit cycle $\omega_{LC}$ must satisfy the following relation [7]

$$N(i\omega_{LC})P(i\omega_{LC})L(i\omega_{LC}) = -1, \quad (26)$$

where $N$ is the describing function [7] of $\Psi$. In such a loop, $\Psi$ may be one of any number of nonlinear elements such as saturation or a relay with a dead zone [7]. The simplest of these elements is an ideal relay with nonlinear amplitude response

$$\Psi(x) = \begin{cases} 1 & \text{if } x \geq 0 \\ -1 & \text{if } x < 0 \end{cases}. \quad (27)$$

If $\Psi$ is chosen as in (27), then by [7] the describing function is positive and real,

$$N_{IR} \in \mathbb{R}_{>0}, \quad (28)$$

where $N_{IR}$ is the describing function of (27). By (26) this implies the rule

$$P(j\omega_{LC})L(i\omega_{LC}) \in \mathbb{R}_{<0}. \quad (29)$$

If $L$ has many modes and natural frequencies, then for some chosen $P$, (29) is satisfied for a set of frequencies

$$S = \{\omega : P(j\omega)L(i\omega) \in \mathbb{R}_{<0}\}, \quad (30)$$

where $S$ is the set of all limit cycle possibilities of the feedback loop. For example, if the response of $L$ at it's the natural frequencies is purely imaginary [22], as is the case with lightly damped general second order systems [22], then

$$L(j\omega_n) \approx -i|L(i\omega_n)|, \quad (31)$$

for natural frequencies $\omega_n$. If one then selects $P = \tilde{P}$ such that

$$\tilde{P}(i\omega) = -i|\tilde{P}(i\omega)| \; \forall \; \omega, \quad (32)$$

then by (31) and (32), all the natural frequencies of $L$ will satisfy the rule (29)

$$L(i\omega_n)\tilde{P}(i\omega_n) \in \mathbb{R}_{<0}, \; S = W, \quad (33)$$

where $W$ is the set of all natural frequencies of $L$ [8]. Therefore, the limit cycle frequency will be one of the natural frequencies of the system. Two common realizations of (32) are an integrator or a negative differentiator $\tilde{P} = \dfrac{1}{s}$ and $\tilde{P} = -s$.

Now enclose the annulus in an AR loop as in Fig.4 with the intention of exciting $\omega_{11}$. Taking (30) and replacing $G_{tr}(\theta,s)$ for $L$, this results in the possible limit cycle set

$$S_{tr} = \{\omega : P(\mathrm{i}\omega)G_{tr}(\theta, \phi_{act}, \mathrm{i}\omega) \in \mathbb{R}_{<0}\} \ . \tag{34}$$

The next step is to select a $P$ such that $\omega_{11}$ will be a limit cycle possibility. That is, we require

$$\omega_{11} \in S_{tr}. \tag{35}$$

If (35) doesn't hold then the AR will not lock to $\omega_{11}$. Now reconsider $G_{tr}(\theta, \mathrm{i}\omega_{11}, \phi_{act})$ so that we may find a $P$ such that (35) is satisfied. Expanding (25), with real $\Gamma_c$ and imaginary $\Gamma_s$ from (18), and assuming $G_{11}(\mathrm{i}\omega_{11})$ is imaginary,

$$G_{tr}(\theta, \phi_{act}, \mathrm{i}\omega_{11}) = -\mathrm{i} \left| \phi_c(\theta) \Gamma_c(\phi_{act}) G_{11}(\mathrm{i}\omega_{11}) \right| \pm \left| \phi_s(\theta) \Gamma_s(\phi_{act}) G_{11}(\mathrm{i}\omega_{11}) \right| \ . \tag{36}$$

It can be seen that unlike $L$ with a phase response dependent on only two parameters $\theta$ and $s$, the phase response of $G_{tr}$ is dependent on three parameters $\theta$, $s$, and $\phi_{act}$. This means for a constant $P$, if one changes $\phi_{act}$, the set (34) will also change and (35) may not be satisfied. Unfortunately, AR and travelling wave control are coupled, and attempting to change the torque level will also change the limit cycle frequency. This is unsatisfactory and instead we require for a constant $P$, that the excitation frequency $\omega_{11}$ always be maintained, regardless of the torque applied. Therefore, It's desired that

$$\frac{\partial \angle G_{tr}(\theta, \phi_{act}, s)}{\partial \phi_{act}} = 0 \ \forall \phi_{act} \ . \tag{37}$$

It's observed in (36) that the contribution from the cosine term is always imaginary and the contribution from the sine term is always real. It appears that the weighting through $\Gamma_c(\phi_{act})$ and $\Gamma_s(\phi_{act})$ and their sum is the mechanism causing the unwanted change in phase response from $\phi_{act}$. Therefore, it's conceivable that canceling either $\eta_c$ or $\eta_s$ from the feedback signal may solve the coupling problem by generating an effective plant (as seen by the feedback loop) with constant phase for all $\phi_{act}$. This can be accomplished with a modal filter [28, 30, 31, 32, 33].

In their paper, Davis et al [8] combine a modal filter (MF) with AR (MFAR) and demonstrated the ability to automatically excite a multi-degree of freedom system at resonance, and at the same time, choose the excitation mode. The MF approach required multiple sensors, and by estimating the mode shapes, a linear combination of the sensors was used as the feedback signal. Because these linear combinations are bi-orthogonal, all modal responses but one were effectively canceled out.

If modal filtering is to be used here, it should first be determined if it is indeed beneficial, and if so, which modal response should be canceled. Referring to (18), it's observed that

$$\frac{\partial \angle \Gamma_c(\phi_{act})}{\partial \phi_{act}} = 0$$
$$\frac{\partial \angle \Gamma_s(\phi_{act})}{\partial \phi_{act}} \neq 0 \quad \forall \phi_{act}. \tag{38}$$

Therefore, for the system $G_{tr}$ operating at a constant frequency $\omega_{11}$, it can also be seen that

$$\frac{\partial \angle \phi_c(\theta) \Gamma_c(\phi_{act}) G_{11}(s)}{\partial \phi_{act}} = 0$$
$$\frac{\partial \angle \phi_s(\theta) \Gamma_s(\phi_{act}) G_{11}(s)}{\partial \phi_{act}} \neq 0 \quad \forall \phi_{act}, \tag{39}$$

and the phase response of the cosine mode is immune to changes in $\phi_{act}$. Therefore, if $\eta_s$ can be canceled from the feedback signal, (37) will be satisfied and the AR loop will be decoupled from traveling wave control.

To employ the multiple-sensor approach for the modal filter as proposed in [8], only two sensors would be required as only two modal responses are present. However, in the case of the annulus only a single sensor is needed. This is because one can simply place the sensor directly over a node of the sine mode so that its response will not be present in the feedback signal. In mathematical terms:

$$\begin{aligned}\phi_c(\theta_c) &= A\cos(\kappa\theta_c) = A \\ \phi_s(\theta_c) &= A\sin(\kappa\theta_c) = 0\end{aligned}, \quad \theta_c = \frac{2\pi N}{\kappa}, \quad N \in \mathbb{Z}, \tag{40}$$

where by (23) $\theta_c$ is measured from the reference actuator. From (40) a modal filter can be realized simply by placing the feedback sensor at location $\theta_c$. Then the effective plant controlled by AR will be

$$G_{tr}(\theta_c, \phi_{act}, s,) = A\Gamma_c(\phi_{act}) G_{11}(s). \tag{41}$$

And the phase response of (41) at the natural frequency is

$$\angle G_{tr}(\theta_c, \phi_{act}, i\omega_{11}) = \angle A\Gamma_c(\phi_{act}) G_{11}(i\omega_{11}) = -\pi/2 \ \forall \phi_{act}. \tag{42}$$

Using a $\tilde{P}$ as in (32) and placing the sensor at $\theta_c$,

$$S_{tr} = \left\{\omega : \tilde{P}(i\omega) G_{tr}(\theta_c, \phi_{act}, i\omega) \in \mathbb{R}_{<0}\right\} = \omega_{11} \ \forall \phi_{act}, \tag{43}$$

and (35) is satisfied. The phase response of the effective plant is constant with respect to $\phi_{act}$, and the AR loop is decoupled from $\phi_{act}$ and traveling wave control.

### 6.1 AR with a bandpass filter

It was demonstrated that by filtering out $\eta_s$ from the feedback signal, that AR can be completely decoupled from travelling wave control. For the truncated system $G_{tr}$, with only one natural frequency this means that the limit cycle will occur at that frequency, i.e. $S_{tr} = \omega_{11} = \omega_{LC}$. However, to analyze the complete behavior of the annulus excited by AR, the complete untruncated system must be considered. Taking the complete response (4) and inserting (18)

$$U(\theta, \phi_{act}, s) = FG_{11}(s)\left[\phi_c(\theta)\Gamma_c(\phi_{act}) + \phi_s(\theta)\Gamma_s(\phi_{act})\right] + F\sum_{\kappa=1 \notin 11}^{\infty} G_\kappa \left[\phi_{c,\kappa}(\theta)\Gamma_{c,\kappa} + \phi_{s,\kappa}(\theta)\Gamma_{s,\kappa}\right]. \tag{44}$$

Dividing by the reference $F$ as in (25) one obtains the complete transfer function of the annulus.

$$G(\theta, \phi_{act}, s) = \frac{U}{F}$$
$$= G_{11}(s)\left[\phi_c(\theta)\Gamma_c(\phi_{act}) + \phi_s(\theta)\Gamma_s(\phi_{act})\right] + \sum_{\kappa=1 \notin 11}^{\infty} G_\kappa(s)\left[\phi_{c,\kappa}(\theta)\Gamma_{c,\kappa} + \phi_{s,\kappa}(\theta)\Gamma_{s,\kappa}\right] \tag{45}$$

Executing the modal filter by choosing sensor location $\theta_c$, the effective plant is becomes

$$G(\theta_c, \phi_{act}, s) = A\Gamma_c(\phi_{act})G_{11}(s) + \sum_{\kappa=1 \notin 11}^{\infty} G_\kappa(s)\left[\phi_{c,\kappa}(\theta_c)\Gamma_{c,\kappa} + \phi_{s,\kappa}(\theta_c)\Gamma_{s,\kappa}\right]. \tag{46}$$

Inserting (45) and (32) (30)

$$S = \{\omega_{11}, \omega_{1c}, \omega_{2c}, ...\} = \{\omega : \tilde{P}(i\omega)G(\theta_c, i\omega)\}, \tag{47}$$

where $\omega_{\kappa c}$ are the other frequencies of $G$ that satisfy (47) at $\theta_c$. At this point it becomes clear that modal filtering only guarantees that $\omega_{11} \in S$ is a limit cycle possibility. Unfortunately, many other frequencies are limit cycle possibilities as well, and by Davis et al [8] the true limit cycle frequency will only satisfy

$$|P(i\omega_{LC})G(\theta_c, \phi_{act}, i\omega_{LC})| = |PG(\theta_c)|_{max}, \quad \omega_{LC} \in W. \tag{48}$$

Currently, there is no guarantee that $\omega_{11}$ will satisfy (48). However, if a bandpass filter is inserted into the AR loop, we may be able to amplify the contribution of $\eta_c$ in the response, and ensure the system can only be excited at $\omega_{11}$. When a bandpass filter $B$ is added to the AR loop, (48) becomes [8, 29]

$$|B(i\omega_{LC})P(i\omega_{LC})G(\theta_c, i\omega_{LC})| = |BPG(\theta_c)|_{max}, \quad \omega_{LC} \in S. \tag{49}$$

And the limit cycle will only occur at the frequency which satisfies (49). A diagram of the AR feedback loop with a bandpass filter can be seen in Fig.5.

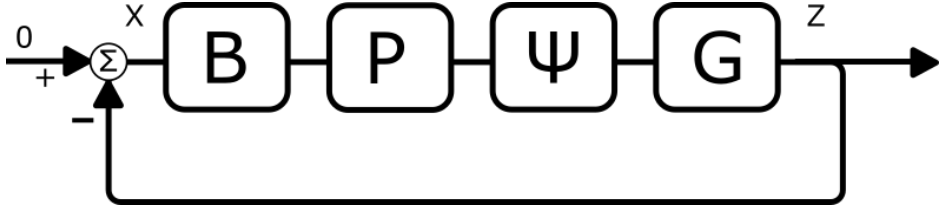

Figure 5. AR loop with bandpass filter

The purpose of $B$ is to attenuate the response of the other modes from the feedback signal so that the right hand side of (49) will be maximized at $\omega_{11}$. Therefore, a useful bandpass filter $B_{\omega_{11}}$ will have a center frequency near $\omega_{11}$ and satisfy

$$\left|B_{\omega_D}(i\omega_{11})\right| \approx \left|B_{\omega_{11}}\right|_{\max} . \tag{50}$$

Just as before, calculate the phase response of $\angle B_{\omega_{11}}(i\omega_{11})G(i\omega_{11})$ and choose $P$ accordingly to ensure $\omega_{11} \in S$. But if $B_{\omega_{11}}$ can be chosen such that its phase response near $\omega_{11}$ is negligible

$$\angle B_{\omega_{11}}(i\omega_{11}) \approx 0 , \tag{51}$$

then $P$ can be realized as in (32), just as if $B_{\omega_{11}}$ was not present. Therefore, $\omega_{11}$ will still be included in the set,

$$\omega_{11} \in S = \left\{\omega : \tilde{P}(i\omega)B_{\omega_{11}}(i\omega)G(\theta_c, i\omega) \in \mathbb{R}_{<0}\right\} , \tag{52}$$

and the solution to (49) is likely $\omega_{11}$.

$$\left|B_{\omega_{11}}(i\omega_{LC})P(i\omega_{LC})G(\theta_c, i\omega_{LC})\right| =$$
$$\left|B_{\omega_D}(i\omega_{11})P(i\omega_{11})G(\theta_c, \phi_{act,} i\omega_{11})\right| = \left|B_{\omega_{11}}PG(\theta_c)\right|_{\max} , \omega_{LC} \in S \tag{53}$$

The limit cycle frequency is likely to be $\omega_{LC} = \omega_{11}$. In the experimental section, an effective $B_{\omega_{11}}$ implemented digitally will be presented.

7. **Open Loop Response Experiment: Procedure**

The purpose of this experimental set was to determine the effectiveness of the annulus model (5) and the modal control scheme as described in (16). Specifically, we want to know if the modal forces behave as in (17) and if the modal response behave as in (19) as $\phi_{act}$ is varied. When exciting the annulus frequency $\omega$, by (19)

$$U_{tr}(\theta_c,\phi_{act},i\omega) = A\eta_c = AFG_{11}(i\omega)\Gamma_c(\phi_{act})$$
$$U_{tr}(\theta_s,\phi_{act},i\omega) = A\eta_s = AFG_{11}(i\omega)\Gamma_s(\phi_{act})$$
, (54)

for antinode locations of the sine mode $\theta_s$. When $\omega$ is constant, $AFG_{11}(i\omega)$ is some complex constant. Therefore, if $\phi_{act}$ is varied but $\omega$ held constant, then the resulting modal response plot should be identical to Fig.2 but multiplied by the complex constant $AFG_{11}(i\omega)$. Additionally, frequency sweeps will be obtained for various constant values of $\phi_{act}$ to determine if the predicted modal response is valid for a range frequencies near $\omega_{11}$. By (23), the choice of reference actuator is arbitrary.

In later experiments, the rotor will be acoustically levitated above the annulus. Therefore, the sensors were placed under the annulus facing up so the rotor would not block them in the future. To measure individual modal responses, the first sensor was placed 1 wavelength away from the reference actuator (some $\theta_c$), and the second placed 1.25 wavelengths away from the reference actuator (some $\theta_s$). The number of integer wavelengths was arbitrary. A diagram of the sensor locations can be seen in Fig.6.

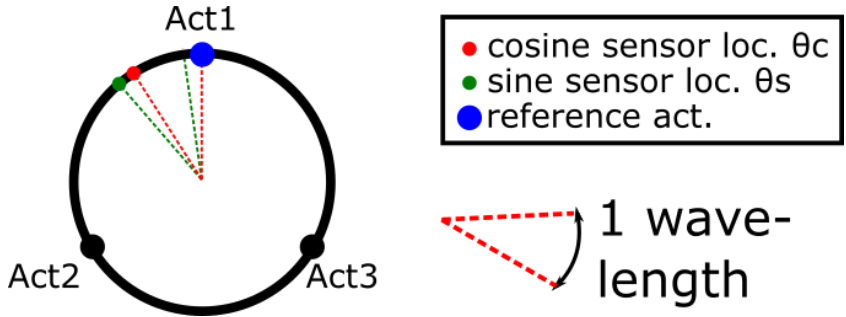

Figure 6. Chosen reference actuator and sensor locations for measuring $A\eta_c$ and $A\eta_s$.

### 7.1 Open Loop Response Experiment: Hardware

To generate the actuation signals, a Nexys 4 FPGA controller [34] capable of sample rates of 1MHz was programmed as a waveform generator. Per (16), the frequency and relative phase of the signals was communicated by a host computer through UART protocol. The outputs of the Nexys 4 are digital, and therefore three square waves from 0 to 3.3 Volts were generated. These signals were passed to a Kemo VBF 40 digital filter. Each channel of this filter was configured for AC coupling, a bandpass filter at center frequency 28kHz, and amplification of 4. This converted each square wave to a sine wave of 6.6 Volts, and no DC offset. Next, the 6.6 Volt waveforms were passed to A.A. Lab Systems x20 voltage amplifiers with a multiplication of 20. This amplified the waveforms to a final amplitude of 132 Volts. These signals were passed to the piezoelectric actuators.

To acquire data, a National Instruments PXIe-6358 Data Acquisition Device with a sample rate of 1 MHz was used. To sense the vibration response, two Keyence LK-H008 with a resolution of $0.005\mu m$ were used. These signals were recorded by the DAQ. Additionally, the 6.6 Volt reference actuator signal was also passed to the DAQ. All data sets were measured over a period of 0.1 seconds. A diagram of this setup can be seen in Fig.7.

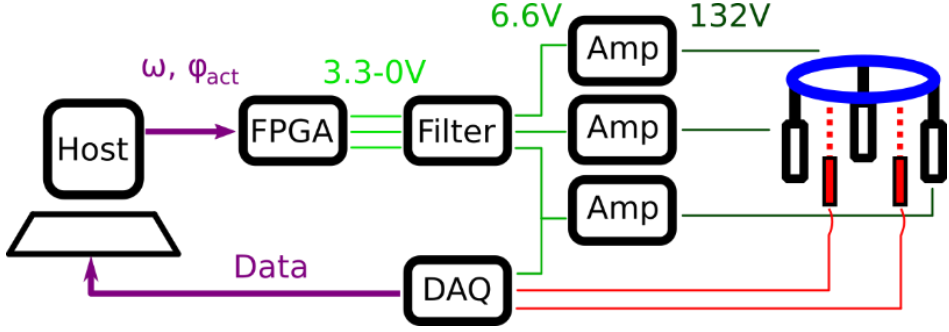

Figure 7. Hardware setup for measuring the modal responses $A\eta_c = AFG_{11}(\mathrm{i}\omega)\Gamma_c(\phi_{act})$ and $A\eta_s = AFG_{11}(\mathrm{i}\omega)\Gamma_s(\phi_{act})$ to changes in $\phi_{act}$

### 7.2 Open Loop Modal Response Experiment: Signal Processing and Results

To generate the complex modal response plots, the reference actuator signal was used as the phase reference, as in (45). To extract this response, the synchronous detection method [35] was used. This required the in-phase and quadrature of the acquired reference actuator signal, and the analytic signal [35] was calculated for this purpose,

$$F_A = Fe^{-as} + \mathrm{i}\tilde{F}e^{-as}, \tag{55}$$

where $F_A$ is the analytic signal of acquired $F$, $\tilde{F}$ is the Hilbert Transform [35] of $F$, $a$ is a time delay introduced by the calculation, and $s$ is the excitation frequency. The amplitude of the reference was calculated as

$$A_F = |F_A|. \tag{56}$$

If we also use the analytic signal of the vibration signals, then the phase shift from the time delay is negated. Therefore, the modal responses were calculated as

$$\begin{aligned} A\eta_c &= \frac{2\overline{\mathrm{Re}(F_A)\mathrm{Re}(S_{cA})}}{A_F} + \mathrm{i}\frac{2\overline{\mathrm{Im}(F_A)\mathrm{Re}(S_{cA})}}{A_F} \\ A\eta_s &= \frac{2\overline{\mathrm{Re}(F_A)\mathrm{Re}(S_{sA})}}{A_F} + \mathrm{i}\frac{2\overline{\mathrm{Im}(F_A)\mathrm{Re}(S_{sA})}}{A_F} \end{aligned}, \tag{57}$$

where $\overline{\bullet}$ is the mean on $\bullet$, $S_c$ and $S_s$ are the measured signals from the cosine and sine sensors, and $S_{cA}$ and $S_{sA}$ are the analytic signals of that data. The resulting modal response plots can be seen in Fig.8.

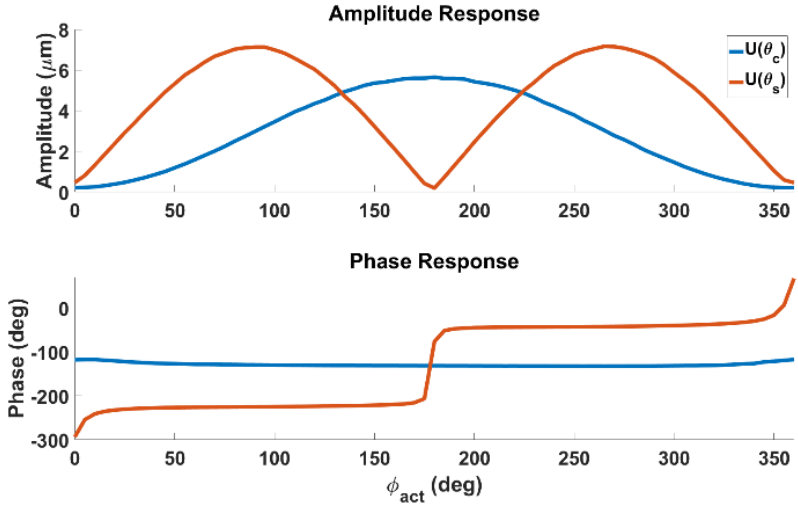

Figure 8. Measured modal response plot $A\eta_c = AFG_{11}(i\omega)\Gamma_c(\phi_{act})$ and $A\eta_s = AFG_{11}(i\omega)\Gamma_s(\phi_{act})$ (54) at excitation frequency 27960 Hz

Clearly the response of the cosine and sine modes is indeed described by (54) and resembles Fig.2 multiplied by a complex constant $AFG_{11}(i\omega)$. Additionally, the frequency response of the doublet modes are plotted below to see if the behavior of (54) holds over a frequency range. This can be seen in Figs.9-11.

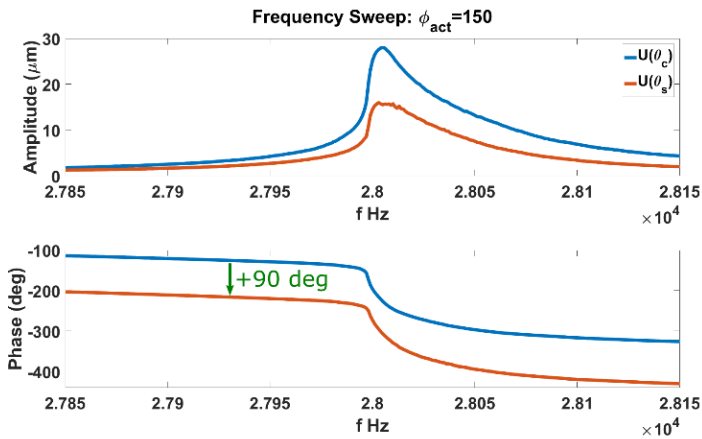

Figure 9. Measured modal frequency sweep of $A\eta_c = AFG_{11}(i\omega)\Gamma_c(\phi_{act})$ and $A\eta_s = AFG_{11}(i\omega)\Gamma_s(\phi_{act})$, $\Delta f = 300 Hz$, $\phi_{act} = 150 \deg$

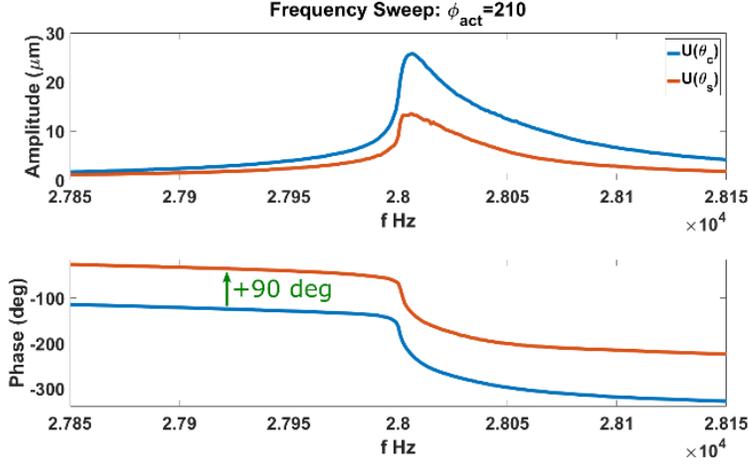

Figure 10. Measured modal frequency sweep of $A\eta_c = AFG_{11}(\mathrm{i}\omega)\Gamma_c(\phi_{act})$ and $A\eta_s = AFG_{11}(\mathrm{i}\omega)\Gamma_s(\phi_{act})$, $\Delta f = 300 Hz$, $\phi_{act} = 210\deg$

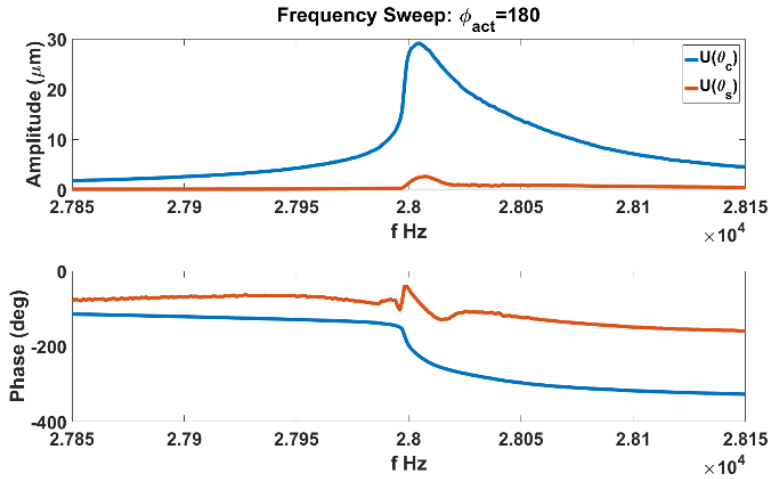

Figure 11. Measured modal frequency sweep of $A\eta_c = AFG_{11}(\mathrm{i}\omega)\Gamma_c(\phi_{act})$ and $A\eta_s = AFG_{11}(\mathrm{i}\omega)\Gamma_s(\phi_{act})$, $\Delta f = 300 Hz$, $\phi_{act} = 180\deg$

It's observed over a $300 Hz$ frequency range that the amplitude response of the annulus is asymmetric and therefore includes some nonlinear effects. But aside from the asymmetry, it does appear that both the amplitude and phase behave per (54), which is dependent on the response of $\Gamma_c$ and $\Gamma_s$. For example, in Fig.9, the response from the sine mode is shifted -90 degrees with respect to the cosine mode, where in Fig.10 the shift is +90 degrees. This is predicted by (18) and (54), i.e. $\angle\left(\Gamma_s(150°)/\Gamma_c(150°)\right) = -\pi/2$, $\angle\left(\Gamma_s(210°)/\Gamma_c(210°)\right) = \pi/2$. Additionally,

$$\left|\Gamma_s(150°)/\Gamma_c(150°)\right| = \left|\Gamma_s(210°)/\Gamma_c(210°)\right| \approx \frac{1}{2},$$

which is indeed observed in Figs.9,10. Finally one can see in Fig.11 that the amplitude of the sine response is nearly zero for all swept frequencies. Again this is in line with (18), i.e. $|\Gamma_s(180°)| = 0$. It's speculated that the small nonzero amplitude response of the sine mode is due to nonlinear behavior of the piezoceramic in the actuator. These results are interpreted as a simultaneous validation of the annulus model (5), and the modal control scheme (16), (17).

## 8. Closed Loop Response Experiment

In this experiment, the modal response plot was generated again, but this time while the annulus was excited by an AR loop as seen in Fig.5. This was to determine if the levitation modes were still controllable while the annulus was automatically excited at its natural frequency. Modal filtering was also applied by using the response of the cosine mode (sensor placed at $\theta_c$ Fig.6) as the feedback signal. By (41) this should decouple AR from changes in $\phi_{act}$ and the plot should resemble Figs.2,8. The control algorithm was programmed on the Nexys 4 FPGA controller board [34].

When designing the bandpass filter $B_{\omega_{11}}$, it was noted in the frequency sweeps (Figs.9-11), that the unloaded natural frequency of the annulus was $\omega_{11} \approx 28kHz$. Therefore, to satisfy (50), the center frequency was chosen to be $28kHz$. To satisfy condition (51), $B_{\omega_{11}}$ was chosen to be two lightly damped biquad filters [36] in series. This gave a very flat phase response in the passband. The passband was chosen to be only 220 Hz wide to ensure no other vibration modes could be accidently excited. This filter was discretized using the bilinear transformation [37], with a time step of 1e-7 seconds ($10MHz$).

$$
\begin{aligned}
B_1 &= \frac{1}{s^2 + 2\zeta(\omega_B - \Delta) + (\omega_B - \Delta)^2} \\
B_2 &= \frac{1}{s^2 + 2\zeta(\omega_B + \Delta) + (\omega_B + \Delta)^2} \\
B_{\omega_B} &= -B_1 B_2
\end{aligned}
\quad , \omega_B = 28 \; kHz, \; \zeta = 9e-5, \; \Delta = 110 \; Hz \quad (58)
$$

A bode plot of these filter in series can be seen in Fig.12.

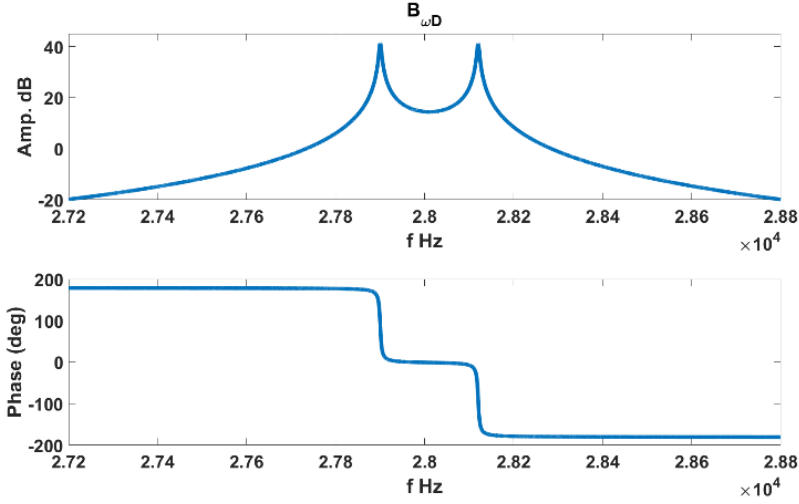

Figure 12. Bandpass filter in Autoresonance loop $B_{\omega_{11}}$ before discretization with extremely flat phase response in the passband $\sim 0\deg$ as in (51).

Next, a $P$ was designed to satisfy (52). In practice, a latency and additional phase shift is introduced by the A/D converter. The sample rate of the Nexys 4's A/D converter was $1MHz$. By the Shift Theorem the A/D converter would introduce a phase delay of [38]

$$\phi_{AD} = \frac{180}{\pi} \angle e^{-i\frac{2\pi 28e3}{1e6}} \Rightarrow -10\deg. \tag{59}$$

Therefore, to satisfy (52), $P$ would need to correct the latency phase shift, in addition to adding a phase shift of $-\pi/2$. A simple solution was to create a quadrature signal within the FPGA algorithm. With this, the sum of the in-phase and quadrature signals with different weights could shift the signal by any arbitrary amount. Because the natural frequency of the annulus was previously known to be $\omega_{11} \approx 28kHz \pm 100Hz$, and since the signal consists of only a single harmonic, a simple numerical integration and scaling by $\omega_{11}$ created an effective quadrature signal, i.e.

$$\tilde{U} \approx -2\pi \frac{28e3}{s}. \tag{60}$$

And to shift the phase of signal $U$ by any amount $\phi$ [38] on the FPGA

$$Ue^{i\phi} = U\cos(\phi) + \tilde{U}\sin(\phi). \tag{61}$$

This technique was also used to shift the phase of the non-reference actuator signals by $\pm\phi_{act}$. A diagram of the combined Autoresonance and traveling wave control algorithm programmed on the Nexys 4 can be seen in Figure 13. As before, $\phi_{act}$ was communicated to the FPGA from the host computer through UART.

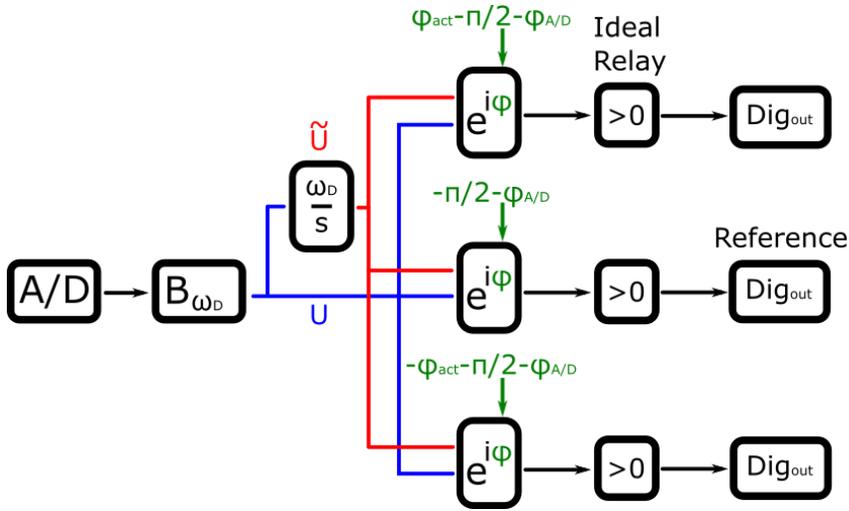

Figure 13. Algorithm programmed on FPGA for combined Autoresonance and travelling wave control. The $e^{i\phi}$ blocks are as described in (61).

The new closed loop hardware setup was similar to the open loop configuration. The only difference being the response from the cosine mode $U(\theta_c)$ was now fed to the FPGA as the feedback signal for AR. Additionally, the frequency was not chosen by the host, as the excitation frequency was always $\omega_{11}(t)$. A diagram of this hardware setup can be seen in Fig.14.

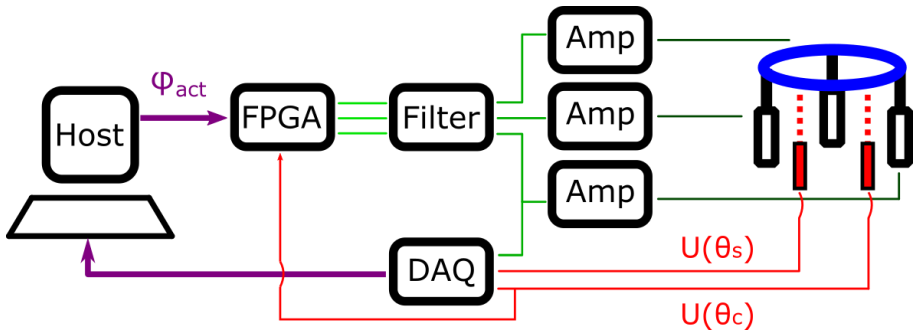

Figure 14. Hardware setup for measuring the vibration response of $A\eta_c = AFG_{11}(i\omega_{11})\Gamma_c(\phi_{act})$ and $A\eta_s = AFG_{11}(i\omega_{11})\Gamma_s(\phi_{act})$ to changes in $\phi_{act}$ while the annulus is excited by Autoresonance feedback loop

Now that the system was automatically excited at $\omega_{11}(t)$, a phase plot was generated by varying $\phi_{act}$ and recording the modal responses $U(\theta_c)$ and $U(\theta_s)$. This can be seen in Fig.15.

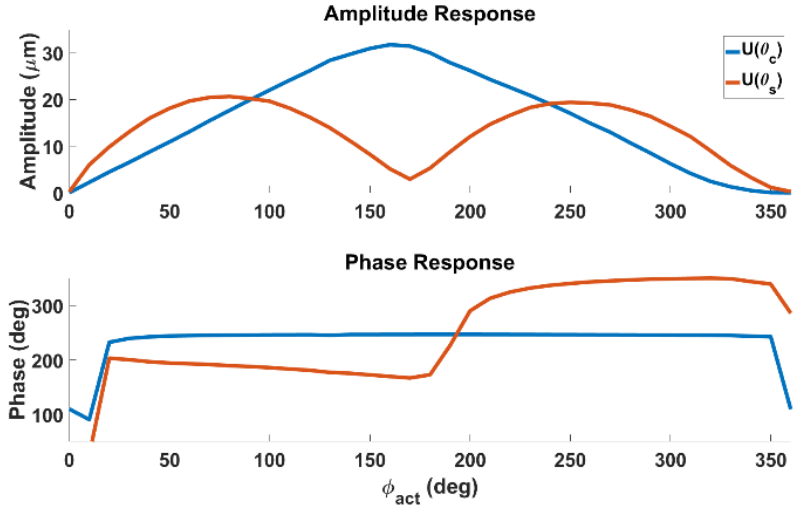

Figure 15. Measured modal response plot of $A\eta_c = AFG_{11}(i\omega_{11})\Gamma_c(\phi_{act})$ and $A\eta_s = AFG_{11}(i\omega_{11})\Gamma_s(\phi_{act})$ while system is automatically excited by AR at frequency $\omega_{11}(t)$

It can be seen in Fig.15 that even though the system was excited by AR (note the higher amplitude compared to Fig.8), the modal response still strongly resembles Fig.2 multiplied by the complex constant $AFG_{11}(i\omega_{11})$, though slightly warped. This warping was likely due to the fact that AR and traveling wave control were not completely decoupled. Additionally, the nonlinearities of the system were clearly amplified near resonance, as was seen by the frequency sweeps. It did appear however, that the modal behavior was still good enough for traveling wave control.

## 9. Acoustic Motor Validation Experiment

The purpose of the final experiments were to demonstrate a functioning acoustic motor. Using an encoder to record the rotation of the levitated rotor, the controllability of the steady state speed and direction could be analyzed. Additionally, by recording the modal vibration responses of the annulus, the amplitude and sign of the travelling wave component $A_T$ could be extracted as described by (9) and compared to the steady state speed of the rotor. This was used to validate the hypothesized proportionality between the traveling wave response of the annulus and the applied torque on the rotor (9).

In this experimental setup, a thin plastic disc of roughly 200 grams was placed on top of the annulus and levitated. This disk acted as the rotor. Because there was nearly no friction between the surfaces, the disk needed to be constrained laterally so that it would not slide off the annulus. A thin rod was connected to the base of the motor. A hole was cut through the center of the disk and the rod inserted through the hole. This acted as a simple axial bearing. A housing was connected to the center of the disk and a magnet for an encoder placed on top. The housing was designed so that it would not touch the rod. To record the angular speed of the disk, a two-phase magnetic encoder with a resolution of 12 bits per rotation was placed above the center of rotor, just above the magnet at the top of the housing. A cut-away of this configuration can be seen in Fig.16.

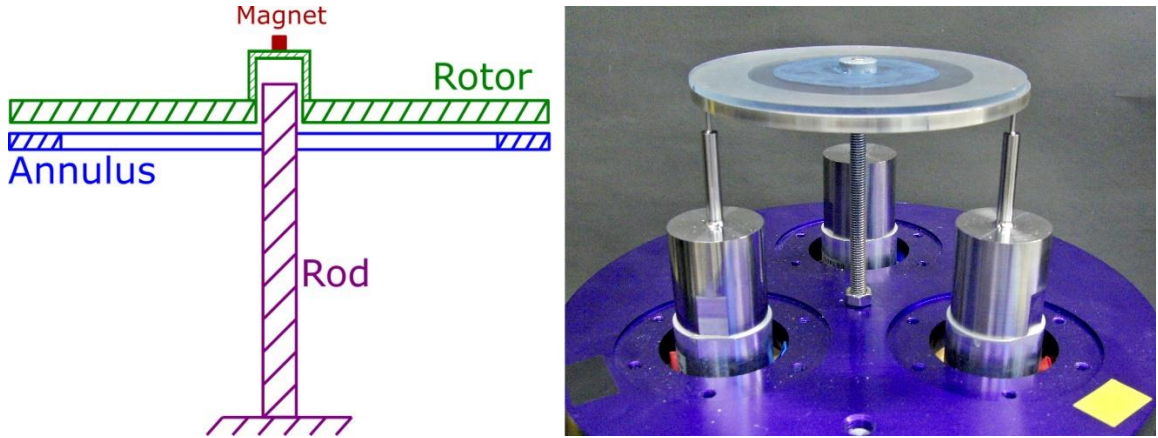

Figure 16. (Left) Cut-away of the rotor configuration, makeshift axial bearing and magnet for two-phase encoder. (Right) Photograph of annulus with actuators, rod and rotor.

To record the steady state speed, the signals from the encoder were recorded by the DAQ. A diagram of the hardware configuration can be seen in Figure 17.

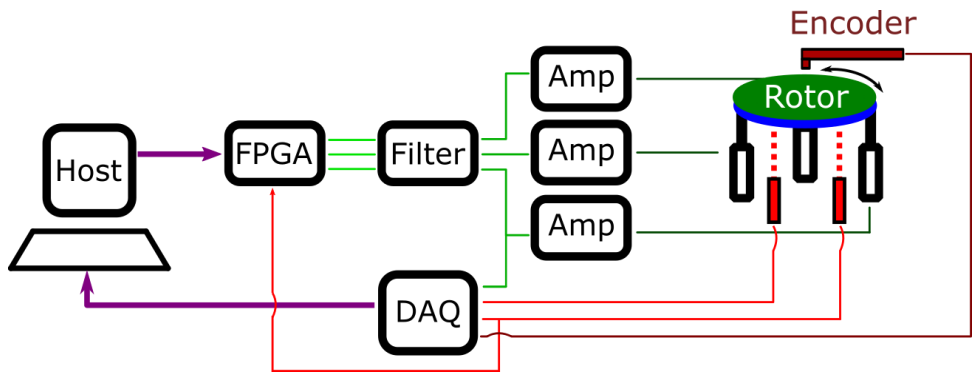

Figure 17. Hardware configuration for validation of the acoustic motor while steady levitation is maintained by Autoresonance feedback loop. Via vibration sensors, $A_T$ was measured and via an encoder steady-state speed of the rotor were measured

In Figs.18,19, two data sets from the acoustic motor experiments are presented. For each set, $\phi_{act}$ was varied from $100$ to $260 \deg$, which was the monotonic region of the applied torque. The left axis represents the amplitude of the traveling wave component $A_T$ which was extracted from the modal vibration data. The right axis shows the steady state speed of the rotor which was calculated from the encoder data.

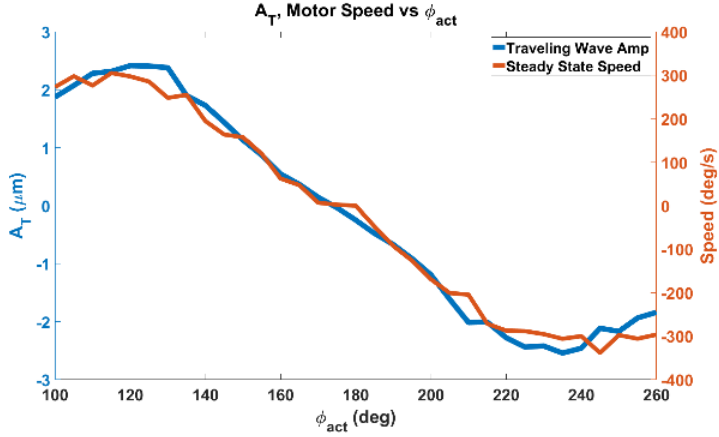

Figure 18. Measured traveling wave component of annulus vibration response $A_T$, and steady state speed of the levitated rotor as $\phi_{act}$ was varied, first data set. The monotonic region is roughly between 120 to 240 degrees, as in Fig.3.

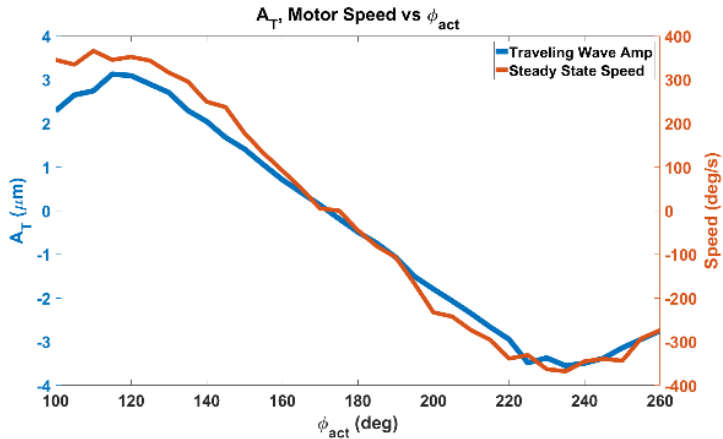

Figure 19. Measured traveling wave component of annulus vibration response $A_T$, and measured steady state speed of the levitated rotor as $\phi_{act}$ was varied, second data set. The monotonic region is roughly between 120 to 240 degrees, as in Fig.3.

From the data sets of Fig.18,19 it's observed that the angular speed and direction of the levitated rotor clearly reacts to $\phi_{act}$ in a predictable and symmetric manner and is very similar to what was seen in Fig.3. For example, one can notice that the monotonic region of the torque roughly resides in the region $120 \leq \phi_{act} \leq 240$, just as in Fig.3. Furthermore, it's observed that the shape of $A_T$ is very similar to the shape of the steady state speed. Therefore it's concluded that $A_T$ is indeed proportional or nearly proportional to the steady state speed. Assuming that the speed is also proportional to the applied torque on the rotor, then (11) is declared validated.

## 10. Conclusion

Though functioning near-field acoustic levitation motors were achieved in the past, these devices were not automatic, as all lacked a resonance tracking feedback loop. This component is essential for stabilizing the levitation of the rotor, and by adding it, the performance and repeatability were greatly increased. Indeed, under AR control, both the ability to acoustically levitate and control the speed of rotation were improved significantly. In addition to positive end results, a good understanding of the modal behavior of the annulus was demonstrated by proving with experiments, that the nearly-axisymmetric structure model of a well-machined annulus is valid. In previous methodologies the operators assumed to know little of their system and preferred a kind of trial-error method for their control scheme. But because it was shown that the annulus behaved as a nearly axisymmetric structure, such an approach is not necessary and a more analytical control scheme is preferred. Additionally, with the measured annulus response and some signal processing it was even shown that one can estimate the rotor's rotational response by measuring the stator's vibration response. Therefore, analysis and experiments of this research clearly demonstrates that an automatic, stable and repeatable near-field acoustic levitation motor is realizable.